\documentclass[lettersize,journal]{IEEEtran}
\usepackage{amsmath,amsfonts}
\usepackage{algorithmic}
\usepackage{algorithm}
\usepackage{array}
\usepackage[caption=false,font=normalsize,labelfont=sf,textfont=sf]{subfig}
\usepackage{textcomp}
\usepackage{stfloats}
\usepackage{url}
\usepackage{verbatim}
\usepackage{graphicx}
\usepackage{cite}
\usepackage[dvipsnames]{xcolor}
\usepackage{tikz}           
\usetikzlibrary{%
    external,               
    calc,                   
    shapes.geometric,       
    arrows,                 
    fit,                    
    trees,                  
    decorations.pathmorphing,
    patterns,
    spy,
    svg.path,
}

\usepgflibrary{%
    plotmarks,              
}

\usepackage{pgfplots}       
\usepgfplotslibrary{%
    external,               
    statistics,             
    groupplots,             
    smithchart,             
    polar,                  
    fillbetween,            
}
\pgfplotsset{compat=newest}

\usepackage{tikzscale}      

\usepackage[%
    european,
  ]{circuitikz}             
  
\usepackage[
      exponent-product={\cdot},       
      per-mode=fraction,              
      detect-all,
  ]{siunitx}

\usepackage{bm}
\usepackage{booktabs}
\usepackage{ifthen}
\usepackage{etoolbox}
\usepackage{mathtools}

\usepackage{hyperref}
\usepackage{scalerel}
\definecolor{orcidlogocol}{HTML}{A6CE39}
\tikzset{
	orcidlogo/.pic={
		\fill[orcidlogocol] svg{M256,128c0,70.7-57.3,128-128,128C57.3,256,0,198.7,0,128C0,57.3,57.3,0,128,0C198.7,0,256,57.3,256,128z};
		\fill[white] svg{M86.3,186.2H70.9V79.1h15.4v48.4V186.2z}
		svg{M108.9,79.1h41.6c39.6,0,57,28.3,57,53.6c0,27.5-21.5,53.6-56.8,53.6h-41.8V79.1z M124.3,172.4h24.5c34.9,0,42.9-26.5,42.9-39.7c0-21.5-13.7-39.7-43.7-39.7h-23.7V172.4z}
		svg{M88.7,56.8c0,5.5-4.5,10.1-10.1,10.1c-5.6,0-10.1-4.6-10.1-10.1c0-5.6,4.5-10.1,10.1-10.1C84.2,46.7,88.7,51.3,88.7,56.8z};
	}
}

\newcommand\orcidicon[1]{\,\textsuperscript{\href{https://orcid.org/#1}{\mbox{\scalerel*{
\tikzexternaldisable
\begin{tikzpicture}[yscale=-1,transform shape]
\pic{orcidlogo};
\end{tikzpicture}
\tikzexternalenable
}{|}}}}}
    
\DeclareSIUnit\dBm{dBm}
\DeclareSIUnit\dBsm{dBsm}
\DeclareSIUnit\dBi{dBi}
\DeclareSIUnit\au{a.u.}
\DeclareSIUnit\rad{rad}
\DeclareSIUnit\GS{GS}
\DeclareSIUnit\ppb{ppb}

\newcommand{\blacklinesolid}{\raisebox{2pt}{\tikz{\draw[-,black,solid,line width = 1.0pt](0,0) -- (4.6mm,0);}}}
\newcommand{\blacklinedashed}{\raisebox{2pt}{\tikz{\draw[-,black,densely dashed,line width = 1.0pt](0,0) -- (4.6mm,0);}}}
\newcommand{\redarrowcaption}{\raisebox{2pt}{\tikz{\draw[-stealth,red,solid,line width = 1.0pt](0,0) -- (4.6mm,0);}}}
\newcommand{\yellowmark}{\raisebox{-0.5pt}{\tikz{\node[yellow,mark size=0.7ex]{\pgfuseplotmark{diamond*}};}}}\newcommand{\bluebox}{\raisebox{-0.5pt}{\tikz{\node[blue,mark size=0.7ex]{\pgfuseplotmark{square*}};}}}

\makeatletter
\def\subsubsection{%
  \@startsection
    {subsubsection}                 
    {3}                             
    {\parindent}                    
    {3.5ex plus 1.5ex minus 1.5ex}  
    {0.7ex plus .5ex minus 0ex}     
    {\normalfont\normalsize\itshape}
}
\makeatother

\begin{document}

\title{UAV-Borne Digital Radar System for \\ Coherent Multistatic SAR Imaging}

\author{Julian Kanz,
Christian Gesell,
Christina Bonfert,
David Werbunat, 
Alexander Grathwohl,
Julian Aguilar,\\
Martin Vossiek, and
Christian Waldschmidt
\thanks{
This work was supported by the Deutsche Forschungsgemeinschaft (DFG, German Research Foundation) under the project GRK 2680-Project-ID 437847244.\\
\indent Julian Kanz, Christian Gesell, Christina Bonfert, David Werbunat, Alexander Grathwohl, Julian Aguilar, and Christian Waldschmidt are with the Institute of Microwave Engineering, Ulm University, 89081 Ulm, Germany (e-mail: julian.kanz@uni-ulm.de).\\
\indent Martin Vossiek is with the Institute of Microwaves and Photonics, Friedrich-Alexander-Universität Erlangen-Nürnberg (FAU), 91058 Erlangen, Germany.\\
\indent This manuscript has been submitted to the IEEE Journal of Microwaves for consideration.}}



\maketitle

\begin{abstract}
Advancements in analog-to-digital converter (ADC) technology have enabled higher sampling rates, making it feasible to adopt digital radar architectures that directly sample the radio-frequency (RF) signal, eliminating the need for analog downconversion.
This digital approach supports greater flexibility in waveform design and signal processing, particularly through the use of digital modulation schemes like orthogonal frequency division multiplexing (OFDM).
This paper presents a digital radar system mounted on an uncrewed aerial vehicle (UAV), which employs OFDM waveforms for coherent multistatic synthetic aperture radar (SAR) imaging in the L-band.
The radar setup features a primary UAV node responsible for signal transmission and monostatic data acquisition, alongside secondary nodes that operate in a receive-only mode.
These secondary nodes capture the radar signal reflected from the scene as well as a direct sidelink signal.
RF signals from both the radar and sidelink paths are sampled and processed offline.
To manage data storage efficiently, a trigger mechanism is employed to record only the relevant portions of the radar signal.
The system maintains coherency in both fast-time and slow-time domains, which is essential for multistatic SAR imaging.
Because the secondary nodes are passive, the system can be easily scaled to accommodate a larger swarm of UAVs.
The paper details the full signal processing workflow for both monostatic and multistatic SAR image formation, including an analysis and correction of synchronization errors that arise from the uncoupled operation of the nodes.
The proposed coherent processing method is validated through static radar measurements, demonstrating coherency achieved by the concept. 
Additionally, a UAV-based bistatic SAR experiment demonstrates the system’s performance by producing high-resolution monostatic, bistatic, and combined multistatic SAR images.
\end{abstract}


\renewcommand{\arraystretch}{1.2}

\section{Introduction}
\label{sec:intro}
Recent advances have led to the successful demonstration of digital radar systems across various application domains~\cite{Sturm2009, Hakobyan2019, Roos2019, Schweizer2021, GirotodeOliveira2022, Moro2024}. 
These systems utilize high-speed digital-to-analog (DAC) and analog-to-digital converters (ADC) to directly generate and sample radio-frequency (RF) baseband signals. 
The use of digital signal generation offers significantly greater flexibility in waveform design compared to analog radar systems. 
Moreover, digital modulation techniques such as orthogonal frequency division multiplexing (OFDM) support joint communication and sensing~\cite{GirotodeOliveira2022, BaqueroBarneto2019, Chen2020}.


Radar networks further enhance imaging performance by capturing data from multiple perspectives~\cite{Werbunat2021}, effectively increasing the aperture and improving angular resolution~\cite{Werbunat2024a}. 
In this context, digital modulation schemes like OFDM can exploit multiplexing strategies such as spectral interleaving to improve system efficiency~\cite{Roos2019}.

However, maintaining synchronization across multiple radar nodes presents a major challenge. 
Particularly synthetic aperture radar (SAR) imaging requires coherency in time, frequency, and phase. 
Various methods have been proposed to ensure coherency, including hardware-based coupling through shared RF signals~\cite{Ash2015, Shin2017} or low-frequency (LF) reference links~\cite{Durr2022, Stefko2022, Merlo2024, Janoudi2024}. 
Additionally, error mitigation techniques that exploit error symmetry have been shown to further improve coherency~\cite{Gottinger2021, Fenske2024}.

An alternative approach is to achieve coherency through post-processing without direct coupling between nodes~\cite{Durr2022, Janoudi2024, Frischen2017, Frischen2020, Werbunat2024}. 
Passive radar systems also avoid the need for synchronization but are constrained by the characteristics of external transmitters, such as fixed bandwidth and limited spatial configuration~\cite{Kulpa2011, Gromek2016, Ulander2017}. 
The feasibility of OFDM-based passive radar has also been demonstrated~\cite{Berger2010}. 
In contrast, radar repeater networks provide a dedicated transmit signal and eliminate the need for synchronization, offering a more controlled alternative to passive systems~\cite{Werbunat2021, Werbunat2024a, Meinecke2019}.

Synchronization becomes particularly challenging when radar nodes are widely spaced and exhibit independent motion, as is often the case in uncrewed aerial vehicle (UAV)-based systems~\cite{Podbregar2024}. 
In recent years, UAV-borne SAR systems have been proposed for a variety of applications, including climate change monitoring~\cite{Ullmann2024}, SAR interferometry~\cite{Mustieles-Perez2024, RuizCarregal2024}, landmine detection~\cite{Schartel2018a}, avalanche victim detection~\cite{Grathwohl2021}, and runway debris identification~\cite{Miccinesi2024}, among others~\cite{Frey2019, Svedin2021, Grathwohl2022, Lopez2022, Bekar2022}.
Radar platforms based on digital radio-frequency system-on-chip (RFSoC) technology have recently been implemented on UAVs~\cite{Koderer2022, Prager2022a, Wasik2023, Moro2024}, with some of it employing OFDM waveforms for enhanced signal flexibility and performance~\cite{Moro2024, Koderer2022}. 
These implementations, however, are monostatic and confined to a single UAV.

In contrast, multistatic SAR configurations offer extended capabilities, such as single-pass interferometry~\cite{Krieger2006} and SAR tomography~\cite{Moreira2015}. 
Additional applications include wind speed estimation~\cite{Garrison2002}, soil moisture retrieval~\cite{Pierdicca2008}, three-dimensional surface displacement measurement~\cite{Pieraccini2018}, and analysis of glacial snow cover~\cite{Stefko2024}.
Bistatic UAV-based SAR has been demonstrated using wireless time synchronization techniques~\cite{Wang2022a, Jeon2024} and radar repeater network concepts~\cite{Grathwohl2023, Kanz2024}. 
Passive SAR imaging with UAV-mounted receivers has also been proposed, relying on external frequency-modulated (FM) broadcast transmitters~\cite{Gabard2021}. 
Additionally, a UAV-based bistatic radar system designed for snow measurement has been introduced, which uses local oscillators derived from Global Navigation Satellite System (GNSS) references for synchronization~\cite{Reyhanigalangashi2024b}.

Advancements in DAC and ADC technology have made it possible to directly generate and sample RF radar signals. Sampling rates of $\SI[per-mode = symbol]{4}{\GS\per\second}$ and above allow for alias-free acquisition of L-band signals~\cite{Farley2018, Peters2022, Michalak2022}, eliminating the need for analog modulation and demodulation stages, as well as precise hardware-based synchronization.

Building on the hardware platforms described in~\cite{Farley2018, Peters2022, Michalak2022}, this work presents a coherent multistatic SAR system that operates without synchronizing reference signals across nodes. 
A UAV-borne digital radar architecture is introduced that enables direct RF signal sampling. 
The system employs a field programmable gate array (FPGA)-based backend coupled with a custom high-speed data interface, allowing for flexible waveform generation, including arbitrary transmit signals.
Radar imaging is conducted using an OFDM waveform, and data acquisition is controlled through a trigger mechanism based on a predefined pseudo-noise (PN) sequence. 
This approach significantly reduces the duty cycle, thereby lowering the average data rate while preserving the essential radar data.

The radar nodes operate independently in both monostatic and multistatic modes, which are executed simultaneously to maximize temporal coherence. 
In the monostatic case, standard OFDM radar techniques are combined with SAR processing. 
The multistatic mode leverages a dual-path transmission of the radar signal, where one of the paths --- the direct sidelink between nodes --- serves as a reference for coherent demodulation.
By utilizing this known reference signal, coherency in time, frequency, and phase is established across nodes. 
Accurate absolute range measurements require precise knowledge of the reference signal’s time-of-flight (ToF). 
However, this requirement aligns with the localization precision already needed for UAV-based SAR image formation. 
As a result, the stringent timing synchronization typically required in conventional systems is effectively replaced by a localization accuracy prerequisite, which is an inherent condition for high-resolution SAR.
The passive nature of the multistatic receive nodes enables straightforward scalability to larger UAV swarms, making the system well-suited for distributed radar networks.

The paper is organized as follows:
Section~\ref{sec:concept} introduces the overall concept of coherent multistatic SAR imaging.
Section~\ref{sec:system} describes the UAV-borne digital radar hardware.
Section~\ref{sec:processing} outlines the signal processing for both monostatic and multistatic SAR modes.
Section~\ref{sec:errors} analyzes the errors arising from uncoupled node operation, utilizing both theoretical models and simulations, and demonstrates that these errors are inherently corrected by the proposed approach.
Finally, Section~\ref{sec:measurements} presents experimental validation, including static radar measurements and a bistatic UAV-based SAR measurement to assess system performance and validate the coherency achieved by the proposed concept.

\section{System Concept}
\label{sec:concept}
Maintaining coherency in time, frequency, and phase is essential for multistatic SAR imaging. 
This section introduces a synchronization strategy that leverages the flexibility of digital radar systems to achieve coherency across a swarm of UAVs. 
Additionally, the implications of this approach on signal design and radar waveform selection are examined.

\subsection{Coherent Multistatic SAR Imaging}
\label{subsec:concept_coherent_multistatic_sar}
Synchronization among spatially distributed nodes in a cooperative radar network presents significant challenges. 
This work proposes an alternative approach, where the nodes are not synchronized through conventional means, i.e., by aligning reference signals used to generate the radar waveform. 
Instead, an identical copy of the radar transmit signal is transmitted via the direct path between the nodes. 
This path is referred to as sidelink throughout the remainder of the text.

\begin{figure}[tb]
	\centering
	\includegraphics[width=82mm]{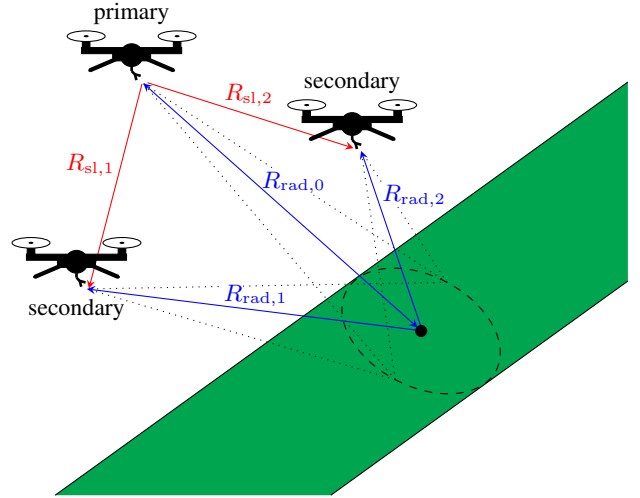}
	\caption{Conceptual llustration of the proposed multistatic SAR approach with a swarm of three UAVs and indicated radar (\textcolor{blue}{rad}) and sidelink (\textcolor{red}{sl}) paths.}
	\label{fig:uav_swarm_concept}
\end{figure}

Fig.~\ref{fig:uav_swarm_concept} illustrates the proposed concept with a swarm of three UAVs. 
Each UAV-based node can function as a transmitter, receiver, or both. 
The multistatic digital SAR system distinguishes between two types of propagation paths: sidelink paths $R_{\mathrm{sl},i}$ and conventional radar paths $R_{\mathrm{rad},i}$. 
Within a given radar cycle, one UAV acts as the primary node, transmitting the radar signal, while the secondary nodes operate in receive-only mode, capturing both radar and sidelink signals.

The system supports two fundamental modes of operation, each with different coherency requirements. 
In the monostatic mode, the transmitting and receiving functions reside on the same node, inherently ensuring coherency since both signal chains share a common reference. 
In the bistatic mode, however, transmission and reception occur on separate, independently operating nodes. 
Without shared references, discrepancies in timing or phase can arise, potentially degrading coherency and impairing SAR image formation.

These coherency challenges are addressed by transmitting the radar signal via two distinct paths: the conventional radar path and the sidelink path. 
The ToF of the sidelink signal can be accurately determined, as precise localization is already required for SAR imaging. 
This sidelink signal serves as a demodulation reference, functionally analogous to the transmit signal in monostatic configurations. 
Since both the radar and sidelink signals are subject to the same system, induced errors effectively cancel out.

This enables simultaneous monostatic and multistatic SAR imaging, maximizing temporal coherence between SAR images. 
Single-pass acquisitions with diverse observation and scattering geometries become feasible. 
Each multistatic receiver operates passively, requiring only to receive the transmit signal via both propagation paths. 
Since only one node transmits at a time, the architecture is inherently scalable to large UAV swarms with multiple radar-equipped platforms.

\subsection{Timing and Localization Requirements}
\label{subsec:concept_requirements}
In conventional time-based synchronization schemes, the required timing precision is defined by the maximum allowable deviation between reference signals. 
In the proposed approach, this stringent timing requirement is instead mapped to the accuracy of the localization data. 
A global time reference is only needed to associate radar measurements with the corresponding localization information.
SAR processing distinguishes between two time domains: fast-time, corresponding to the time within a single radar pulse, and slow-time, referring to the duration of the entire acquisition along single radar measurements. 
Conventional synchronization methods require precision in fast-time to avoid distortion in the SAR signal. 
In contrast, the proposed method ensures that both radar and reference signals are derived from the same clock at each node, requiring synchronization only in slow-time. 
As a result, full coherency in both time domains is achieved with significantly relaxed timing requirements several orders of magnitude lower than in time-based methods.

In the following, an exemplary comparison of the required timing precision of the proposed approach with conventional approaches is presented. 
Assuming a maximum acceptable one-way range error of $R_\mathrm{max}$\,$=$\,$\lambda/10$\,$\approx$\,$\SI{2}{\centi\meter}$, based on the parameters listed in Table~\ref{tab:radar_parameters}, the corresponding allowable timing offset between the transmit and receive nodes is
\begin{equation}
\delta t_\mathrm{max,sync} = 2 \delta R_\mathrm{max}/c_0 \approx \SI{133}{\pico\second}.\label{eq:timing_accuracy_conventional_sync}
\end{equation}
If the timing accuracy falls below the required threshold, distortions are introduced into the radar measurements, which serve as the input for SAR image formation. 
In the proposed synchronization approach, however, each OFDM symbol maintains inherent fast-time coherency, as the radar signal is evaluated relative to the sidelink reference. 
For coherent bistatic SAR processing, the only remaining offset to be corrected is the sidelink path delay caused by the spatial separation between UAVs. 
This delay is compensated by utilizing localization data.
Therefore, the timing accuracy requirement is reduced to ensuring correct temporal alignment between the radar measurements and their corresponding localization information. 
As a result, the necessary timing precision is several orders of magnitude less stringent than that required in conventional time-synchronized systems.

Two primary constraints remain regarding the allowable maximum timing offset.
First, to enable joint processing across multiple receive nodes, their respective measurements must be temporally aligned. 
Therefore, the timing error must remain below the pulse repetition interval $T_\mathrm{pri}$.
Accordingly, the maximum permissible timing offset is
\begin{align}
\delta t_\mathrm{max,pri} &= T_\mathrm{pri} = 1/f_\mathrm{prf} = \SI{10}{\milli\second}\label{eq:timing_accuracy_prf}\\
\intertext{
for the pulse repetition frequency listed in Table~\ref{tab:radar_parameters}. 
A potentially more restrictive limitation arises from the use of time-stamped localization data, which are referenced to the local time base of each node.
Any timing offset translates into a spatial localization error due to the motion of the UAV.
An estimate of this limitation can be expressed as
}
\delta t_\mathrm{max,loc} &= 2 \delta R_\mathrm{max}/v_\mathrm{max} = \SI{4}{\milli\second}\label{eq:timing_accuracy_velocity}
\end{align}
for an assumed maximum velocity $v_\mathrm{max}$\,$=$\,$\SI{10}{\meter\per\second}$.

In the worst case of a bistatic geometry, the required localization accuracy may be reduced to half of that needed for monostatic SAR.
In practical scenarios, however, localization errors at the transmit and receive nodes often do not accumulate linearly and may partially or completely cancel each other.
Additionally, position offsets along the sidelink direction that affect both nodes in the same direction do not impact the sidelink path length.
As a result, the overall localization accuracy requirement is approximately equivalent to that of a monostatic SAR system involving a single UAV.

\subsection{Signal Design}
\label{subsec:concept_signal_design}
The transmit signal is digitally generated and then converted to an analog RF waveform via a DAC. 
This allows for the design of a signal frame that can be tailored to meet specific application requirements. 
As the platform moves along its flight path, this signal frame is continuously repeated to enable synthetic aperture processing.

\begin{figure}[h]
	\centering
	\includegraphics[width=80mm]{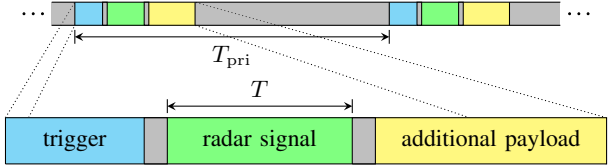}
	\caption{Illustrated signal frame as part of a measurement sequence including trigger and radar parts as well as additional payload.}
	\label{fig:signal_design}
\end{figure}

The primary component of the signal frame is the waveform utilized for radar imaging. 
However, the frame can also include supplementary elements such as a trigger sequence or custom data payloads. 
An example of a repeated signal frame is illustrated in Fig.~\ref{fig:signal_design}, which also highlights the pulse repetition interval (PRI) $T_\mathrm{pri}$ and the duration of the radar signal $T$.

Transmitting a predefined trigger sequence known to the receiver nodes enables precise control over data acquisition. 
Recording is initiated only when the desired signal is detected, ensuring that only relevant portions of the data are stored. 
This approach significantly lowers the average data rate and reduces the burden on the data storage interface. 
The resulting mean data rate can be expressed as
\begin{align}
C
&= f_\mathrm{s} \cdot \nu \cdot \gamma,\label{eq:data_rate}\\
\intertext{where $f_\mathrm{s}$ is denotes sampling frequency, $\nu$ the ADC resolution in bits, and}
\gamma 
&= \frac{T}{T_\mathrm{pri}}\label{eq:duty_cycle}
\end{align}
the duty cycle. 
Here, $T$ represents the duration of the radar signal to be stored, while $T_\mathrm{pri}$ denotes the PRI at which the signal frame is transmitted periodically, see Fig.~\ref{fig:signal_design}.

The signal frame may also include additional payload components, such as communication signals or sequences designed to detect and characterize errors resulting from the absence of hardware-based synchronization between nodes. 
Furthermore, encoding measurement frames can facilitate the association of monostatic and bistatic data without requiring a coarsely synchronized time reference.

\subsection{Radar Waveform}
\label{subsec:concept_radar_waveform}
The radar imaging waveform is defined by selecting an appropriate transmit sequence. 
Since all radar processing is performed digitally in post-processing, modifying the waveform does not require any hardware changes.

While range resolution depends solely on the available bandwidth and not on the specific waveform used, OFDM offers distinct advantages over frequency-modulated continuous-wave (FMCW) or chirp sequence waveforms.
Firstly, OFDM enables the integration of communication and sensing functions by allowing coded communication signals to be repurposed for radar imaging~\cite{BaqueroBarneto2019}. 
Secondly, the effective bandwidth of the OFDM signal is equal to the full transmitted bandwidth, unlike chirp-based signals. 
For chirps of duration $T$, the effective bandwidth is reduced because of the maximum ToF $\Delta t_\mathrm{max}$ expected from any scatterer and is given by~\cite{Fink2015}
\begin{align}
\label{eq:effective_bandwidth}
B_\mathrm{eff} = \frac{T-\Delta t_\mathrm{max}}{T} B.
\end{align}
This limitation becomes particularly relevant in long-range scenarios or when utilizing short chirp durations $T$, where the reduction in effective bandwidth can significantly impact performance. 
In the proposed concept, shorter signal durations $T$ are beneficial, as both the duty cycle~(\ref{eq:duty_cycle}) and the resulting data rate~(\ref{eq:data_rate}) scale proportionally with $T$. 
Given these considerations, OFDM is selected as the radar waveform in this work.


\section{Digital Radar System}
\label{sec:system}
This section introduces the UAV-borne radar system that implements the concept described in Section~\ref{sec:concept}. 
Signal generation and acquisition --- both for the transmit and receive case --- are handled by the digital backend, whereas the frontend serves as the interface between the digital domain and the transmitted or received RF waveforms.

\subsection{Backend}
\label{subsec:system_backend}
The digital radar system is built around a Zynq UltraScale+ RFSoC platform, which is based on an XCZU47DR FPGA together with high-speed DACs and ADCs~\cite{Xilinx2023}. 
The DAC supports sampling rates of $\SI[per-mode = symbol]{9.85}{\GS\per\second}$, while the ADC achieves $\SI[per-mode = symbol]{5}{\GS\per\second}$, enabling direct synthesis and alias-free sampling of RF signals up to  $\SI{2.5}{\giga\hertz}$. 

\begin{figure}[t]
	\centering
	\includegraphics[]{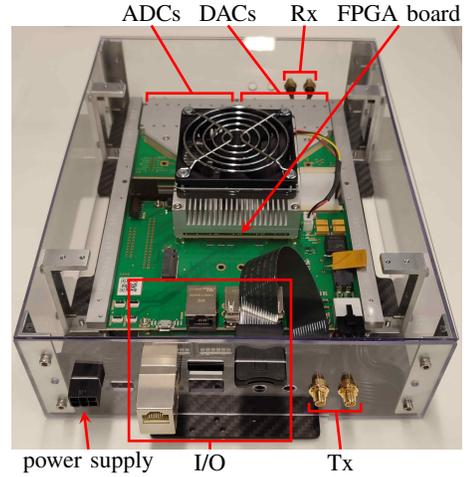}
	\caption{Photograph of the RFSoC-based backend as employed in the proposed digital radar system.}
	\label{fig:photo_system}
\end{figure}

The system provides eight 14-bit channels for both transmitting and receiving. 
In the configuration utilized in this work, two IQ channels are employed: one for receiving the radar signal and another for the sidelink signal. 
An overview of the RFSoC-based backend is shown in Fig.~\ref{fig:photo_system}. 
A coarse time alignment is established via Ethernet utilizing a GNSS-based time reference to support rough pre-synchronization.

To enable local storage of received data, the backend includes a high-speed interface connecting the FPGA to solid-state drives (SSDs), which serves as an integral part of the system architecture. 
The maximum data rate $C_\mathrm{max}$ is constrained by the throughput capabilities of the SSD.
This interface is also employed for signal transmission. 
The desired transmit waveform is read from the SSD, digitally upconverted in real time, and then converted to the analog domain by the DAC.
	
\subsection{Frontend}
\label{subsec:system_frontend}
When operating at frequencies below half the ADC sampling rate, analog upconversion is not required. 
As a result, the analog frontend is limited to essential components such as amplifiers, filters, and antennas.

\begin{figure}[tb]
	\centering
	\includegraphics[width=85mm]{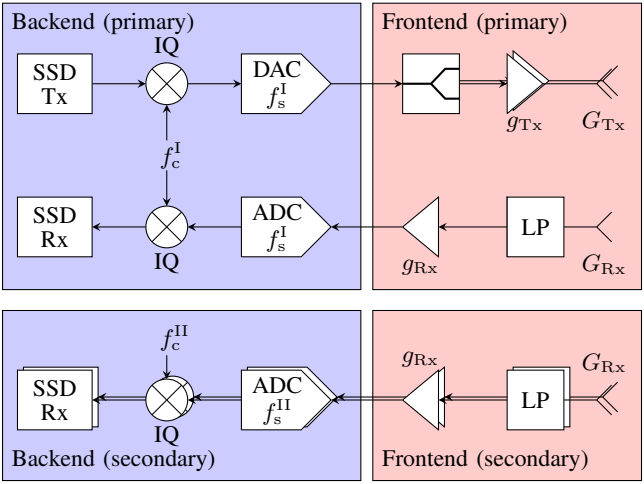}
	\caption{Block diagram illustrating the backend and frontend system components of the primary~(top) and secondary~(bottom) nodes.}
	\label{fig:bsb_system}
\end{figure}

A schematic overview of the system’s backend and frontend architecture is shown in Fig.~\ref{fig:bsb_system}. 
The upper section depicts the monostatic signal paths within the primary node, while the lower section illustrates the bistatic receive paths at a secondary node. 
Only a single transmit channel is used, since the radar and sidelink signals share the same transmit waveform, which is split in the analog domain. 
These signals are transmitted simultaneously and received by the bistatic secondary node. 
In contrast, the primary monostatic node processes only the radar return signal.
The influence of varying carrier frequencies $f_\mathrm{c}^\mathrm{I,II}$ and sampling frequencies $f_\mathrm{s}^\mathrm{I,II}$ between nodes is analyzed in Section~\ref{sec:errors}. 

For the radar path, TEM horn antennas are employed that are described in~\cite{Burr2018a}, while broadband Vivaldi antennas are utilized for the sidelink path.
The gain of the TEM horn antennas is $\SI{6}{dBi}$, while the gain of the Vivaldi antennas is $\SI{8}{dBi}$. 
The beamwidths are $\theta_\mathrm{az}$\,$=$\,$50^\circ$ in azimuth and $\theta_\mathrm{el}$\,$=$\,$60^\circ$ in elevation for the TEM horn antenna, while vertical polarization was utilized.  
For the sidelink Vivaldi antenna, the beamwidths are $\theta_\mathrm{E}$\,$=$\,$55^\circ$ in E plane and $\theta_\mathrm{H}$\,$=$\,$110^\circ$ in H plane. 
The sidelink antennas are installed at a $45^\circ$ angle relative to the ground plane, as illustrated in Fig.~\ref{fig:photo_uav_system}.

\subsection{UAV Platform and Localization}
\label{subsec:system_uav}
\begin{figure}[tb]
	\centering
	\includegraphics[]{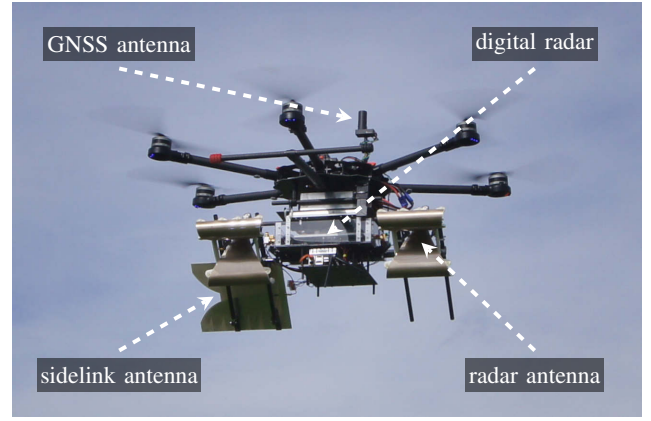}
	\caption{Photograph of the UAV-borne digital radar system.}
	\label{fig:photo_uav_system}
\end{figure}

The digital radar system is integrated into a multicopter platform capable of carrying a payload of up to $\SI{5}{\kilogram}$. 
An overview of the complete UAV-borne system is provided in Fig.~\ref{fig:photo_uav_system}. 
The flight control system is based on a Pixhawk controller, and raw localization data are obtained from an onboard GNSS receiver. 
To enhance positioning accuracy, real-time kinematic (RTK) GNSS is combined with data from an inertial measurement unit (IMU), enabling centimeter-level precision with the approach described in~\cite{Baehnemann2022}. 
Additionally, the GNSS receiver supplies a coarse time reference, which is utilized to associate radar measurements with corresponding localization data for SAR processing.

\section{Signal Processing}
\label{sec:processing}
This section outlines the signal processing workflow for both monostatic and bistatic SAR imaging. 
An OFDM waveform is employed as radar signal, leveraging the advantages discussed in Section~II-\ref{subsec:concept_radar_waveform}.

\begin{figure}[tb]
	\centering
	\includegraphics[width=68mm]{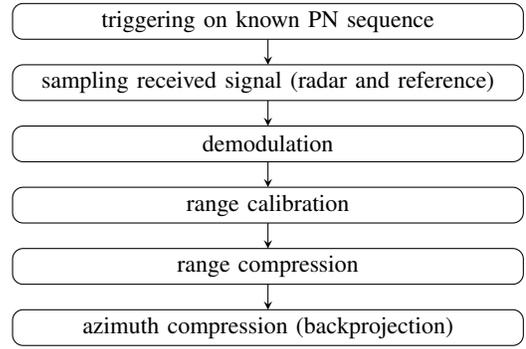}
	\caption{Signal processing workflow for the proposed coherent multistatic SAR concept.}
	\label{fig:flow_chart}
\end{figure}

A flow chart illustrating the complete processing chain is presented in Fig.~\ref{fig:flow_chart}. 
The subsequent sections provide a detailed explanation of each processing step, including the procedures required to achieve coherency and to generate a coherent multistatic SAR image.

\subsection{Signal Model}
\label{subsec:signal_processing_model}
In time-domain, the complex baseband signal of a single OFDM symbol with $N$ subcarriers is given by
\begin{align}
x_\mathrm{b}(t) 
&= \sum_{n=0}^{N-1} d_n \mathrm{e}^{\mathrm{j} 2\pi n \Delta f t} \mathrm{rect} \left( \frac{t-T}{T} \right),\label{eq:signal_ofdm_tx_bb}\\
\intertext{where $n$ is the subcarrier index, $d_n$ the code symbol, $\Delta f$ the subcarrier spacing, and $T$ the symbol duration~\cite{Sturm2009a}. 
A cyclic prefix is applied to the signal but omitted from the equations for clarity.
The bandwidth of the signal is determined by}
B
&= N \Delta f.\label{eq:signal_ofdm_bandwidth}\\
\intertext{The baseband signal $x_\mathrm{b}$ defined in~(\ref{eq:signal_ofdm_tx_bb}) is upconverted utilizing a continuous-wave carrier at frequency $f_\mathrm{c}$ yielding the complex RF signal}
x(t) 
&= \sum_{n=0}^{N-1} d_n \mathrm{e}^{\mathrm{j} 2\pi (n \Delta f + f_\mathrm{c}) t}.\label{eq:signal_ofdm_tx_rf}\\
\intertext{The $\mathrm{rect}$ term is omitted in~(\ref{eq:signal_ofdm_tx_rf}) and in subsequent expressions for clarity.
The transmitted signal is reflected by each target and subsequently received by both the monostatic and multistatic nodes. 
The received signal corresponsing to the $q$th target is given by}
y_q(t) 
&= x(t-\Delta t_q), \quad \text{for}\ q=1,\ldots,Q.\label{eq:signal_ofdm_rx_rf_tar}\\
\intertext{It represents a time-delayed replica of the transmit signal, where the delay corresponds to the ToF}
\Delta t_q
&= \frac{R_{\mathrm{Tx},q}+R_{\mathrm{Rx},q}}{c_0}.\label{eq:tof_tar}
\intertext{Variations in signal amplitude and the relative motion between transmit and receive nodes are neglected in the processing, as their effects can be represented by an additional factor.
The sidelink signal received at a multistatic node is expressed as}
y_\mathrm{sl}(t) 
&= x(t-\Delta t_\mathrm{sl}),\label{eq:signal_ofdm_rx_rf_sl}\\
\intertext{where}
\Delta t_\mathrm{sl}
&= \frac{R_\mathrm{sl}}{c_0}\label{eq:tof_sl}
\end{align}
is the ToF.
The radar path ToF~(\ref{eq:tof_tar}) is valid for a generalized multistatic geometry.
In the special case of a monostatic configuration, the following simplification applies:
\begin{equation}
\label{eq:monostatic_ranges}
R_{\mathrm{Rx},q} 
\overset{\mathrm{mono}}{=} R_{\mathrm{Tx},q}.
\end{equation}
Frequency shifts due to relative motion are neglected, as the resulting Doppler frequencies are several orders of magnitude lower than the carrier frequency in the application under consideration.
For this reason, only a single OFDM symbol is transmitted per measurement, as velocity estimation is not within the scope of this work.

\subsection{Trigger with PN Sequence}
\label{subsec:signal_processing_trigger}
Signal acquisition at the receiver is initiated by a trigger derived from a known pseudo-noise (PN) sequence. 
This sequence is transmitted via the sidelink as a preamble to the radar signal, as illustrated in Fig.~\ref{fig:signal_design}. 
On the FPGA, the received signal is correlated with the known reference, and data recording is triggered when a predefined threshold is exceeded.
As a result, only the relevant radar portion of the received signal is stored locally, effectively excluding unrelated data. 
This recorded segment is then utilized in the subsequent signal processing steps.

\subsection{Sampling of the Received Signal}
\label{subsec:signal_processing_sampling}
The received radar signals given in~(\ref{eq:signal_ofdm_rx_rf_tar}) and~(\ref{eq:signal_ofdm_rx_rf_sl}) are sampled at a rate of $f_\mathrm{s}$ and the resulting samples are stored locally.
The corresponding discrete-time representation of the received signal from the $q$th target is
\begin{align}
y_q[k] 
&= y_q \left( \frac{k}{f_\mathrm{s}} \right) 
= \sum_{n=0}^{N-1} d_n \mathrm{e}^{\mathrm{j} 2\pi (n \Delta f + f_\mathrm{c}) \left( \frac{k}{f_\mathrm{s}} -\Delta t_q \right)}.\label{eq:signal_ofdm_rx_sampled}\\
\intertext{These sampled data serve as the input for coherent post-processing and subsequent SAR image formation.
The known carrier component is eliminated by multiplying the signal with a complex exponential, yielding the corresponding baseband representation}
y_{\mathrm{b},q}[k] 
&= y_q[k] \mathrm{e}^{-\mathrm{j} 2\pi f_\mathrm{c} \frac{k}{f_\mathrm{s}}} \nonumber \\
&= \sum_{n=0}^{N-1} d_n \mathrm{e}^{\mathrm{j} 2\pi n \Delta f \frac{k}{f_\mathrm{s}}} \mathrm{e}^{-\mathrm{j} 2\pi (n \Delta f + f_\mathrm{c}) \Delta t_q}.\label{eq:signal_ofdm_rx_bb}
\end{align}

\subsection{Demodulation}
\label{subsec:signal_processing_demodulation}
A Discrete Fourier Transform (DFT) is applied to the baseband received signal in~(\ref{eq:signal_ofdm_rx_bb}) to extract the received OFDM symbol
\begin{align}
D_{\mathrm{Rx},q}[n]
&= \mathrm{DFT} \left( y_{\mathrm{b},q}[k] \right)_n \nonumber \\
&= \frac{1}{N} \sum_{k=0}^{N-1} y_{\mathrm{b},q}[k] \mathrm{e}^{-\mathrm{j} 2\pi n \Delta f \frac{k}{f_\mathrm{s}}}.\label{eq:signal_ofdm_rx_dft}\\
\intertext{The known transmitted symbols are compensated by applying an element-wise or Hadamard product $\odot$, resulting in data}
D_{n,q}
&= D_{\mathrm{Rx},q} \odot \left( D_{\mathrm{Tx},q} \right)^\ast \nonumber \\
&= \mathrm{e}^{-\mathrm{j} 2\pi (n \Delta f + f_\mathrm{c}) \Delta t_q}\label{eq:signal_ofdm_rx_spectral_division}
\end{align}
that depend on the ToF of the $q$th target $\Delta t_q$. 

\subsection{Range Calibration}
\label{subsec:signal_processing_range_cal}
As outlined in Section~II-\ref{subsec:concept_coherent_multistatic_sar}, the ToF of the target echoes is measured relative to the sidelink distance $R_\mathrm{sl}$. 
To retrieve accurate geometric range information, additional range offsets caused by internal system delays must also be corrected.

For monostatic configurations, the measured ToF
\begin{align}
\Delta t_\mathrm{meas} 
&= \Delta t_\mathrm{geom} + \Delta t_\mathrm{cal}\label{eq:measured_tof}\\
\intertext{corresponds to the sum of the geometric ToF $\Delta t_\mathrm{geom}$ associated with the actual range to the target, and a calibration offset $\Delta t_\mathrm{cal}$, arising from the radar-specific delay $R_\mathrm{cal}$.
This offset introduces an additional phase term in the signal, which must be corrected during processing.
The phase term is estimated by}
D_{\mathrm{cal},n}
&= \mathrm{e}^{-\mathrm{j} 2\pi (n \Delta f + f_\mathrm{c}) \Delta t_\mathrm{cal}},\label{eq:signal_ofdm_rx_spectral_division_cal}\\
\intertext{where}
\Delta t_{\mathrm{cal}}
&= \frac{\hat{R}_\mathrm{cal}}{c_0}\label{eq:tof_cal}\\
\intertext{represents the additional time delay introduced by the estimated radar offset $\hat{R}_\mathrm{cal}$.
This value is obtained through a calibration measurement.
The offset is compensated by}
D_{\mathrm{mono},n,q}
&= D_{n,q} \oslash D_{\mathrm{cal},n} \nonumber \\
&= \mathrm{e}^{-\mathrm{j} 2\pi (n \Delta f + f_\mathrm{c}) \Delta t_{\mathrm{mono},q}},\label{eq:signal_ofdm_rx_spectral_division_cal2}\\
\intertext{where $\oslash$ denotes the Hadamard division. 
The resulting monostatic ToF can then be expressed as}
\Delta t_{\mathrm{mono},q}
&= \frac{2R_{\mathrm{Tx},q} + R_\mathrm{cal} - \hat{R}_\mathrm{cal}}{c_0}
\approx \frac{2R_{\mathrm{Tx},q}}{c_0}.\label{eq:tof_mono}\\
\intertext{For multistatic signals, the same processing step is required.
However, to extract accurate range information, the radar data must be corrected using the sidelink reference signal}
D_{\mathrm{bi},n,q}
&= D_{n,q} \oslash D_{n,\mathrm{sl}} \nonumber \\
&= \mathrm{e}^{-\mathrm{j} 2\pi (n \Delta f + f_\mathrm{c}) \Delta t_{\mathrm{bi},q}}.\label{eq:signal_ofdm_rx_spectral_division_sl}\\
\intertext{The resulting bistatic ToF relative to the sidelink path length is given by}
\Delta t_{\mathrm{bi},q}
&= \frac{R_{\mathrm{Tx},q}+R_{\mathrm{Rx},q}-R_{\mathrm{sl}}}{c_0}.\label{eq:tof_rel}
\end{align}

\subsection{Range Compression}
\label{subsec:signal_processing_rc}
The ToF-dependent data are converted into range profiles by applying an Inverse Discrete Fourier Transform (IDFT) to~(\ref{eq:signal_ofdm_rx_spectral_division_cal2}) and~(\ref{eq:signal_ofdm_rx_spectral_division_sl})
\begin{align}
r_{\mathrm{mono},q}[\rho]
&= \mathrm{IDFT}[D_{\mathrm{mono},n,q}]
= \sum_{n=0}^{N-1} D_{\mathrm{mono},n,q} \mathrm{e}^{\mathrm{j} 2\pi \frac{n \rho}{N}}\label{eq:signal_ofdm_rx_rc}\\
\intertext{for the monostatic and bistatic processing, respectively. 
In the following, only the monostatic case is considered. 
The bistatic range compression follows an analogous procedure with a different indexing scheme.
The resulting range profile is given by}
r_{\mathrm{mono},q}[\rho]
&= \mathrm{e}^{-\mathrm{j} 2\pi f_\mathrm{c} \Delta t_q} \mathrm{sinc} \left( \rho - \rho_{\mathrm{mono},q} \right)\label{eq:signal_ofdm_rx_rc_si}\\
\intertext{with}
\mathrm{sinc}(\rho) 
&= \frac{\sin(\pi \rho)}{\pi \rho},\label{eq:sinc}\\
\intertext{
where $\rho$ denotes the range cell index.
A peak appears at the range}
\rho_{\mathrm{mono},q}
&= \lfloor B \Delta t_{\mathrm{mono},q} \rfloor, \quad \text{for}\ q=1,\ldots,Q\,,\label{eq:signal_ofdm_rx_range_profile_max}
\end{align}
which corresponds to the true geometric range provided that all system-induced offsets have been properly compensated. 

\subsection{Azimuth Compression by Backprojection}
\label{subsec:signal_processing_sar}
Multiple radar acquisitions are conducted along the flight trajectory to synthesize a larger aperture for SAR imaging.
For each measurement $m$, the processing steps outlined in Sections~IV-\ref{subsec:signal_processing_sampling} through \ref{subsec:signal_processing_rc} are applied. 
The resulting range profiles $r_m[\rho]$ are then coherently combined using time-domain backprojection. 
Unlike frequency-domain techniques, this method does not require prior assumptions about the acquisition geometry, making it well-suited for arbitrary UAV flight paths.

A pixel grid is defined over the region of interest for SAR image reconstruction. 
For every pixel and corresponding measurement, an expected phase $\phi_\mathrm{e}$ is computed and subtracted from the measured phase. 
The pixel value
\begin{align}
A(\mathbf{x_0}) 
&= \sum_{m=0}^{M-1} r_m[\rho_m] \mathrm{e}^{-\mathrm{j} \phi_\mathrm{e} (\Delta t_m)}\label{eq:signal_bp}\\
\intertext{is computed by summing the phase-corrected range measurements $r_m[\rho_m]$, where $\rho_m$ denotes the range cell associated with the distance between the radar and the pixel at position $\mathbf{x_0}$. 
The summation is performed over all $M$ radar acquisitions along the flight trajectory~\cite{Zaugg2015}.
The expected phase is determined by}
\phi_\mathrm{e} (\Delta t_m) 
&= - 2\pi f_\mathrm{c} \Delta t_m\,,\label{eq:signal_bp_expectied_phase}
\end{align}
where $\Delta t_m$ represents the ToF corresponding to the bistatic path from the transmitter to the pixel and then to the receiver for measurement $m$. 

Time-domain backprojection for bistatic SAR imaging has been comprehensively described by Grathwohl et al.~\cite{Grathwohl2023}.
The processing steps following range compression in this work are based on that approach, with the exception of the different phase hypothesis for the OFDM waveform, which is defined here according to~(\ref{eq:signal_bp_expectied_phase}).


\section{Incoherency-Induced Errors and Corrections}
\label{sec:errors}
The processing described in the previous section assumes ideal conditions. 
However, in practical multistatic radar systems, discrepancies between the hardware of the transmit and receive nodes can introduce various errors. 
These include deviations in sampling frequencies, carrier frequencies, and time references. 
Additionally, inaccuracies in the localization data can affect the processing chain and degrade the quality of the resulting SAR images.

This section investigates the impact of these error sources through simulation, analyzing their influence on both the range-compressed data and the final SAR images. 
For each type of error, corresponding compensation strategies are discussed. 
All simulation results presented here reflect data prior to the application of~(\ref{eq:signal_ofdm_rx_spectral_division_sl}). 

\begin{table}
\begin{center}
\caption{SAR Simulation Parameters}
\label{tab:sim_parameters}
\begin{tabular}{| l | c | r |}
\hline
Parameter & Symbol & Value\\
\hline
carrier frequency & $f_\mathrm{c}$ & $\SI{1.2}{\giga\hertz}$\\
number of subcarriers & $N$ & $4096$\\
subcarrier spacing & $\Delta f$ & $\SI{100}{\kilo\hertz}$\\
bandwidth & $B$ & $\SI{409.6}{\mega\hertz}$\\
symbol duration & $T$ & $\SI{12.5}{\micro\second}$\\
cyclic prefix duration & $T_\mathrm{cp}$ & $\SI{3.125}{\micro\second}$\\
measurement rate & $f_\mathrm{prf}$ & $\SI{100}{\hertz}$\\
DAC\,/\,ADC sampling rate & $f_s$ & $\SI[per-mode = symbol]{1.024}{\GS\per\second}$\\
\hline
flight altitude above ground & $h$ & $\SI{10}{\meter}$\\
ground-range distance & $r_\mathrm{g}$ & $5$\,$-$\,$\SI{25}{\meter}$\\
trajectory length & $L_\mathrm{sa}$ & $\SI{30}{\meter}$\\
platform speed & $v$ & $\SI[per-mode = symbol]{1}{\meter\per\second}$\\
\hline
theoretical ground-range resolution & $\Delta r_{\mathrm{g}}$ & $35$\,$-$\,$\SI{73}{\centi\meter}$\\
theoretical cross-range resolution & $\Delta r_{\mathrm{c}}$ & $\SI{9}{\centi\meter}$\\
\hline 
\end{tabular}
\end{center}
\end{table}

\subsection{Simulation Parameters}
\label{subsec:errors_sim}
\begin{figure}[tb]
	\centering
	\includegraphics[width=70mm,height=50mm]{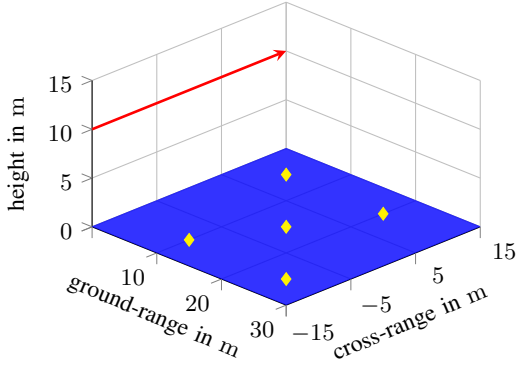}
	\caption{Visualization of the simulation setup containing five point-like targets (\protect\yellowmark) on a pixel grid (\protect\bluebox) and a linear trajectory (\protect\redarrowcaption).}
	\label{fig:sim_scene}
\end{figure}

SAR simulations are conducted utilizing the processing chain outlined in Section~\ref{sec:processing}. 
The simulation parameters, including those related to the OFDM radar waveform and scene geometry, are summarized in Table~\ref{tab:sim_parameters}. 
The resolutions listed in the table are $\SI{3}{\decibel}$\,-\,resolutions. 
A linear flight path is assumed, and the simulated scene consists of five point-like scatterers. 
To isolate the impact of signal processing errors from geometric effects inherent to bistatic configurations, the transmitter and receiver are assumed to be located at the same position. 
The simulation setup is depicted in Fig.~\ref{fig:sim_scene}.

\begin{figure}[tb]
	\centering
	\includegraphics[]{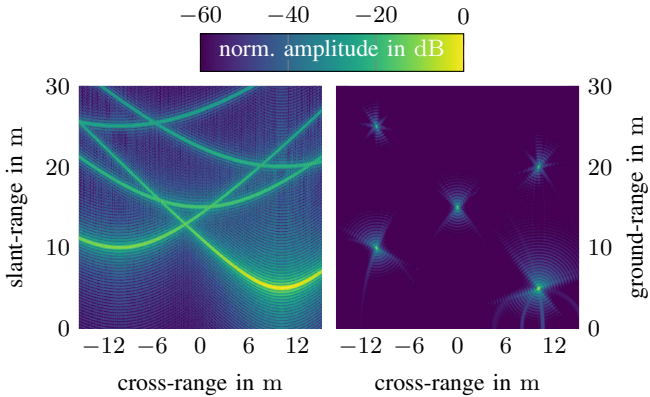}
	\caption{Simulation results of the scene in Fig.~\ref{fig:sim_scene} including range compression (left) and SAR image (right).}
	\label{fig:sim_no_errors}
\end{figure}

The processing chain is first validated by a simulation in which no discrepancies exist between the transmit and receive systems.
The left panel of Fig.~\ref{fig:sim_no_errors} presents the resulting range-compressed signal along the cross-range dimension. 
Following range compression, hyperbolic range profiles are observed for each target along the cross-range direction. 
These patterns arise naturally from the varying slant-range as the platform moves. 
Sidelobes appear in the slant-range dimension due to the $\mathrm{sinc}$-shaped response of the range cells, as described in~(\ref{eq:signal_ofdm_rx_rc_si}). 
Additionally, the amplitude diminishes with increasing slant-range, consistent with free-space path loss.

The corresponding SAR image obtained after azimuth compression is shown in the right panel of Fig.~\ref{fig:sim_no_errors}. 
All targets are correctly focused at their respective ground-range and cross-range positions.

\subsection{Sampling Frequency Offset}
\label{subsec:errors_sfo}
The first error source considered is a mismatch between the sampling frequencies of the transmit and receive nodes, commonly referred to as a sampling frequency offset (SFO).
For generality, it is assumed that the transmitter sampling frequency is  $f_\mathrm{s,Tx}$\,$=$\,$f_\mathrm{s}$, while the receiver sampling rate is defined as $f_\mathrm{s,Rx}$\,$=$\,$\delta_\mathrm{s} f_\mathrm{s}$. 
Accordingly, the SFO is given by
%
\begin{figure}[tb]
	\centering
	\includegraphics[]{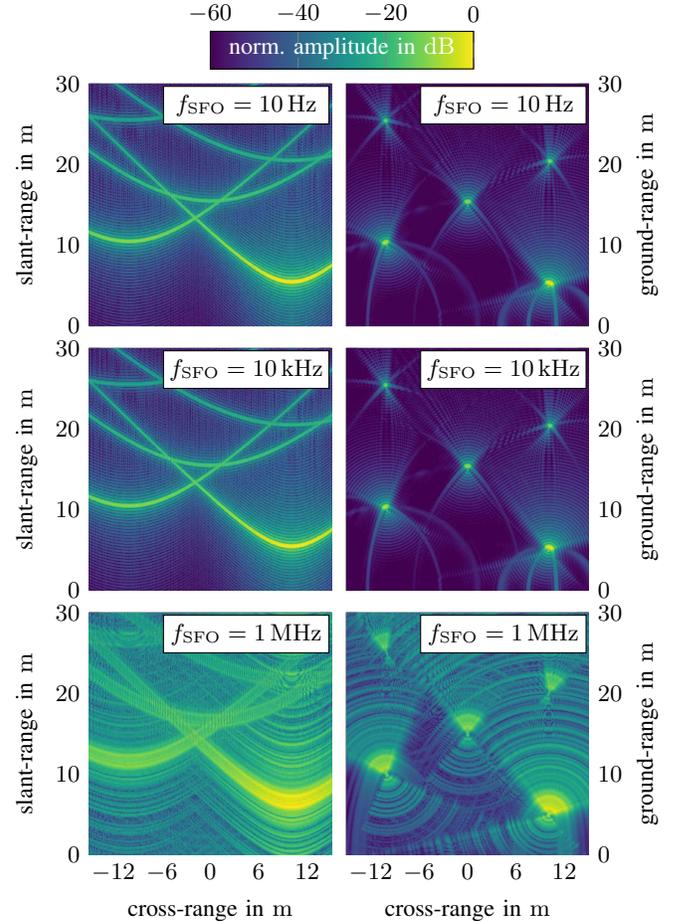}
	\caption{Simulation results of the scene in Fig.~\ref{fig:sim_scene} including range compression (left) and SAR image (right) for different SFO values.}
	\label{fig:sim_sfo}
\end{figure}
%
\begin{align}
f_\mathrm{SFO} 
&= f_\mathrm{s,Tx} - f_\mathrm{s,Rx} 
= (1-\delta_\mathrm{s})f_\mathrm{s}\,.\label{eq:sfo_def}\\
\intertext{
Here, $\delta_\mathrm{s}$ denotes the relative deviation in sampling frequency.
A non-zero SFO results in temporal scaling of the received signal relative to the transmitted waveform, that can either act as compression or stretching of the signal.
This distortion is characterized in~(\ref{eq:signal_ofdm_rx_sampled}). 
To ensure that the impact of the SFO on range compression remains negligible, a condition based on the Nyquist sampling criterion can be established~\cite{Aguilar2024}. 
When the following condition is satisfied, the effects of the SFO can be safely ignored:}
\delta_\mathrm{s} 
&> \frac{1}{1+\frac{1}{BT}}\,.\label{eq:sfo_criterion}
\end{align}
As the time-bandwidth product $BT$ increases, the requirements on sampling frequency accuracy become more stringent.
Given the simulation parameters provided in Table~\ref{tab:sim_parameters}, the maximum allowable SFO is $f_\mathrm{SFO,max}$\,$=$\,$\SI{200}{\kilo\hertz}$, corresponding to a required reference frequency stability of at least $\SI{100}{ppm}$. 
If the actual SFO exceeds this threshold, a dedicated signal component can be transmitted to estimate the offset. 
The resulting distortion can then be corrected through resampling techniques~\cite{Aguilar2024}.

Fig.~\ref{fig:sim_sfo} presents simulation results for various SFO values. 
When the offset surpasses the calculated limit, the range-compressed signal exhibits significant errors, which manifest as strong artifacts in the processed SAR image.

\subsection{Carrier Frequency Offset}
\label{subsec:errors_cfo}
Next, the impact of a carrier frequency offset (CFO) on range compression and SAR image formation is analyzed.
Analogous to the SFO case, the CFO is modeled as
%
\begin{figure}[tb]
	\centering
	\includegraphics[]{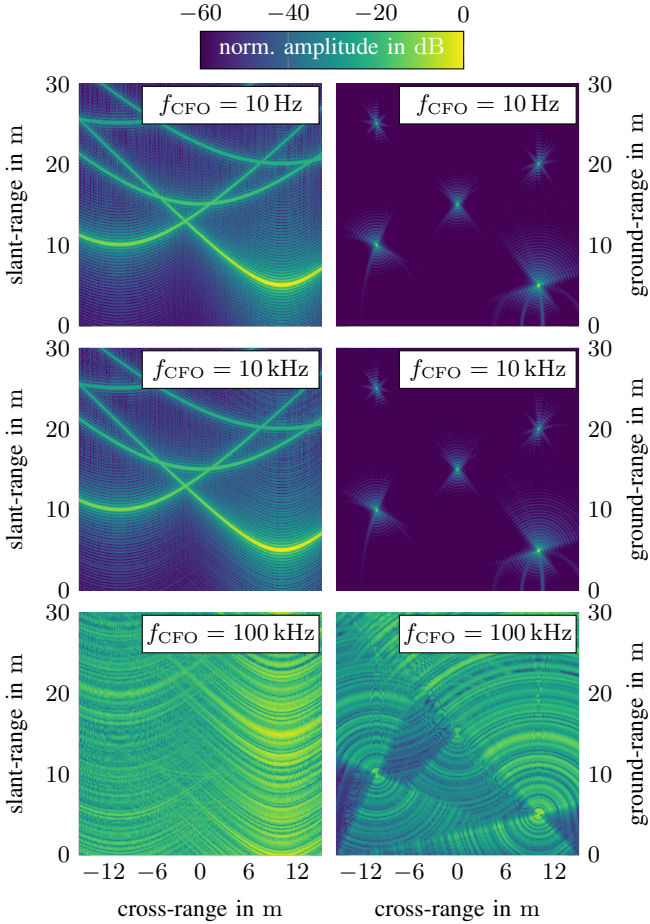}
	\caption{Simulation results of the scene in Fig.~\ref{fig:sim_scene} including range compression (left) and SAR image (right) for different CFO values.}
	\label{fig:sim_cfo}
\end{figure}
%
\begin{align}
f_\mathrm{CFO} 
&= f_\mathrm{c,Tx} - f_\mathrm{c,Rx} 
= (1-\delta_\mathrm{c})f_\mathrm{c}\,.\label{eq:cfo_def}\\
\intertext{A CFO introduces a residual frequency shift of $f_\mathrm{CFO}$, as the high-frequency carrier component in~(\ref{eq:signal_ofdm_rx_bb}) is not fully suppressed.
This residual shift causes intercarrier interference, degrading signal quality.
A criterion for the maximum tolerable CFO can be derived similarly to the case of SFO.
As noted in~\cite{Aguilar2024}, the impact of CFO can be considered negligible if}
\delta_\mathrm{c} 
&> 1-\frac{\Delta f}{10 f_\mathrm{c}}\label{eq:cfo_criterion}
\end{align}
is satisfied. 
Based on the simulation parameters listed in Table~\ref{tab:sim_parameters}, the maximum tolerable CFO is $f_\mathrm{CFO,max}$\,$=$\,$\SI{10}{\kilo\hertz}$. 
As with an SFO, exceeding this threshold leads to distortion in the range-compressed signal, ultimately degrading SAR image quality.
This behavior is demonstrated in the simulation results shown in Fig.~\ref{fig:sim_cfo}, which illustrate the effects of various CFO values.
At the critical value of  $f_\mathrm{CFO}$\,$=$\,$\SI{10}{\kilo\hertz}$, minor deviations are observable, but the SAR image remains largely intact, 
In contrast, significant image degradation occurs in the case with $f_\mathrm{CFO}$\,$=$\,$\SI{100}{\kilo\hertz}$. 

The residual oscillation introduced by the CFO in~(\ref{eq:signal_ofdm_rx_bb}) affects both the radar and sidelink signals identically, as they share the same time reference and sampling index.
As a result, the CFO-induced phase variation can be estimated from the sidelink signal and subsequently compensated for in the radar signal. 

\subsection{Carrier Phase Error}
\label{subsec:errors_cpe}
In SAR imaging, individual measurements $m$ are separated both temporally and spatially.
As a result, slow-time effects arising from changes across multiple acquisitions add to the fast-time distortions within a single measurement, such as those caused by SFO and CFO.

\begin{figure}[tb]
	\centering
	\includegraphics[]{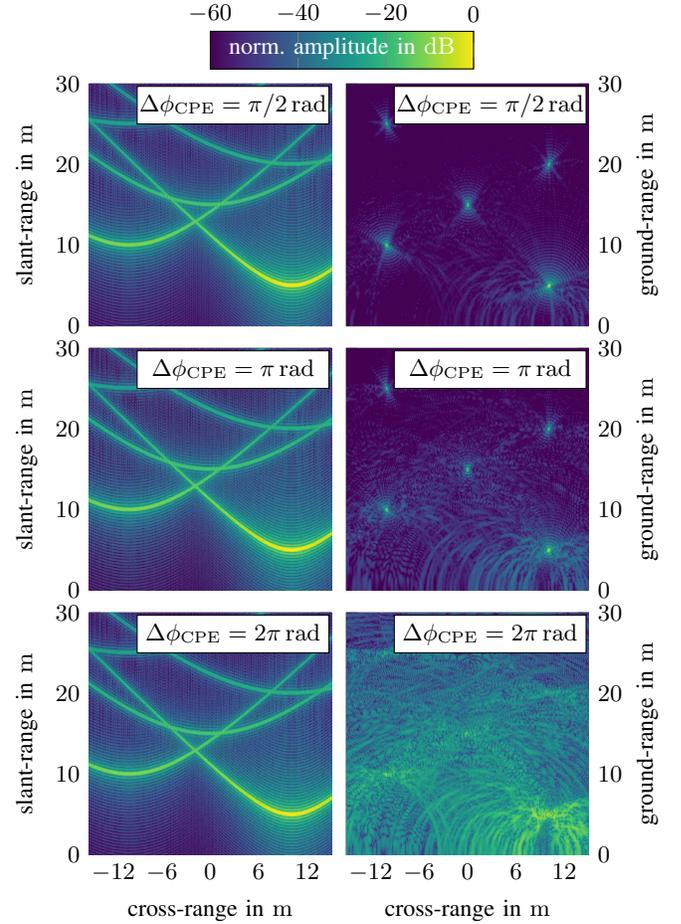}
	\caption{Simulation results of the scene in Fig.~\ref{fig:sim_scene} including range compression (left) and SAR image (right) for different CPE values.}
	\label{fig:sim_cpo}
\end{figure}

The first slow-time effect examined is the carrier phase error (CPE), which arises from a linear phase drift over slow-time due to a CFO.
To simulate this effect, a random phase offset is introduced for each measurement $m$.
The error is modeled as a complex exponential term, applied as a multiplicative factor to the spectral division
\begin{align}
D_{\mathrm{CPE},n,m} 
&= D_{n,m} \mathrm{e}^{\mathrm{j} \phi_{\mathrm{CPE},m}},\label{eq:opo_def}\\
\intertext{where $\phi_{\mathrm{CPE},m}$ represents the carrier phase error associated with measurement $m$.
This phase distortion directly translates into a phase error in the range-compressed signal }
r_{\mathrm{CPE},m}[\rho_m] 
&= r_{m}[\rho_m] \mathrm{e}^{\mathrm{j} \phi_{\mathrm{CPE},m}}.\label{eq:cpo_rc}
\end{align}
Since the range-compressed signal in~(\ref{eq:cpo_rc}) serves as the input to the backprojection algorithm in~(\ref{eq:signal_bp}), any CPE invalidates the expected phase hypothesis.
As a result, azimuth compression becomes erroneous, preventing the generation of a focused SAR image.

Fig.~\ref{fig:sim_cpo} presents simulation results demonstrating the impact of random CPE values across individual measurements.
For each plot, the maximum phase deviation $\Delta \phi_\mathrm{CPE}$ is indicated, with the CPE modeled as a uniformly distributed random variable in the range $-\Delta \phi_\mathrm{CPE}$ to $\Delta \phi_\mathrm{CPE}$. 
As predicted by~(\ref{eq:cpo_rc}), the magnitude of the range-compressed signal remains unaffected. 
However, the quality of the SAR images deteriorates as the CPE increases.

Because the phase error introduced by the CPE is identical in both the radar and sidelink signals, it is inherently compensated when ~(\ref{eq:signal_ofdm_rx_spectral_division_sl}) is applied.
Specifically, the Hadamard division cancels all CPE-related phase components.
This compensation not only addresses CFO-induced phase drift but also mitigates random phase fluctuations, such as those caused by phase noise.

\subsection{Timing Offset}
\label{subsec:errors_to}
Since the transmit and receive nodes operate independently and are not hardware-synchronized, each relies on its own timing reference.
As a result, timing offsets (TOs) may arise between nodes, potentially degrading the quality of the SAR image.
The TO associated with the $m$th radar measurement is defined as
\begin{align}
T_{\mathrm{TO},m}
&= t_{0,\mathrm{Tx}} - t_{0,\mathrm{Rx}}.\label{eq:to_def}\\
\intertext{Here, $t_0$ denotes the start time of the radar signal. 
The timing offset introduces an additional phase term in the spectral division~\cite{Aguilar2024}}
D_{\mathrm{TO},n,m} 
&= D_{n,m} \mathrm{e}^{\mathrm{j} 2\pi (n \Delta f + f_\mathrm{c}) T_{\mathrm{TO},m}}.\label{eq:to_spectral_division}\\
\intertext{A timing offset effectively behaves as an additional ToF, resulting in a corresponding range shift.
Consequently, the range-compressed signal becomes}
r_{\mathrm{TO},m}[\rho_m] 
&= r_{m}[\rho_m+\Delta \rho_{\mathrm{TO},m}] \mathrm{e}^{\mathrm{j} \phi_{\mathrm{TO},m}},\label{eq:to_rc}\\
\intertext{where}
\Delta \rho_{\mathrm{TO},m}
&= B T_{\mathrm{TO},m}\label{eq:to_range_offset}\\
\intertext{is the range offset caused by the TO. 
Furthermore, a phase shift}
\phi_{\mathrm{TO},m} 
&= 2\pi f_\mathrm{c} T_{\mathrm{TO},m}\label{eq:to_phase_offset}
\end{align}
occurs. 
Unlike its impact on range-Doppler processing, as discussed in~\cite{Aguilar2024}, the constraints on $T_\mathrm{TO}$ are different in SAR systems, where the phase consistency across measurements is critical.
In addition to introducing a fast-time range shift, the timing offset also causes a phase error as expressed in ~(\ref{eq:to_phase_offset}), which can vary across measurements $m$.
This results in a slow-time phase disturbance analogous to the CPE previously described.

\begin{figure}[tb]
	\centering
	\includegraphics[]{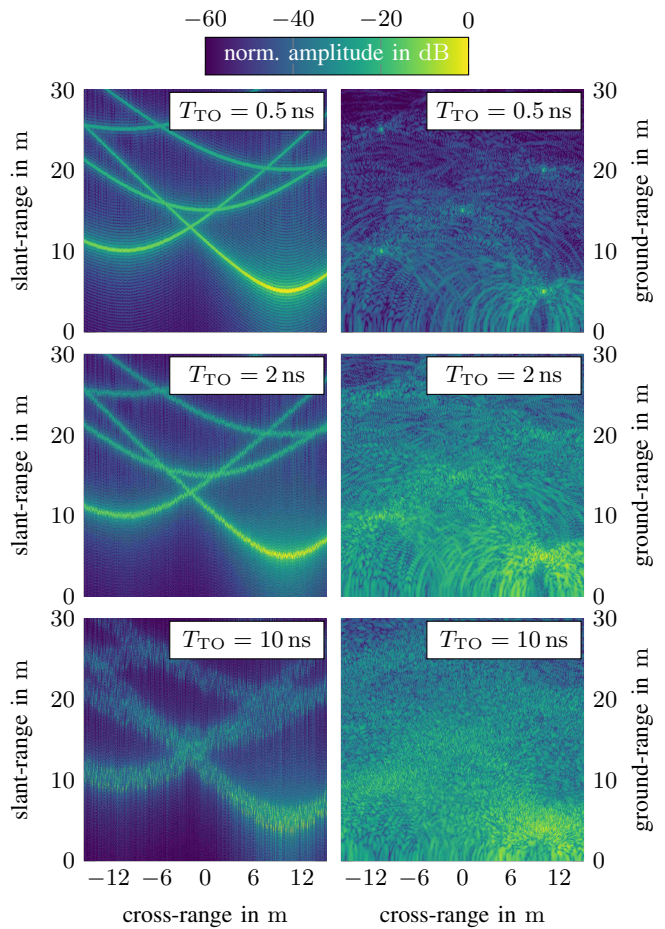}
	\caption{Simulation results of the scene in Fig.~\ref{fig:sim_scene} including range compression (left) and SAR image (right) for different TO values.}
	\label{fig:sim_to}
\end{figure}

The impact of an uncompensated TO on both range-compressed data and SAR image is illustrated in Fig.~\ref{fig:sim_to}.
Random TOs are simulated with maximum values $T_\mathrm{TO}$ indicated in the figure, following the approach used in Section~V-\ref{subsec:errors_cpe}.
Each TO introduces a range shift in the corresponding measurement $M$, resulting in an incorrect phase assumption in~(\ref{eq:signal_bp}) and, consequently, a degraded SAR image.

Even in cases where the absolute range-compressed data remain unaffected, as seen with $T_\mathrm{TO}$\,$=$\,$\SI{0.5}{\nano\second}$ in Fig.~\ref{fig:sim_to}, SAR image quality deteriorates if the timing offset is not corrected.
Therefore, compensating for a TO is essential during processing.
As with a CPE, this correction is achieved by applying~(\ref{eq:signal_ofdm_rx_spectral_division_sl}).

\subsection{Localization Errors}
\label{subsec:errors_loc}
Accurate localization data are essential for successful SAR image formation.
As indicated by~(\ref{eq:signal_bp_expectied_phase}), precise knowledge of the ToF $\Delta t_m$ is required to calculate the expected phase, which is critical for the performance of the backprojection algorithm.

Any offset in the UAV’s three-dimensional position introduces phase errors in the expected phase calculation. 
This issue is not unique to the proposed multistatic concept but also applies to monostatic UAV-based SAR systems.
However, in the multistatic case, additional errors may arise from inaccuracies in estimating the sidelink path length, further affecting phase accuracy when processing the multistatic radar signals.

To simulate these effects, random position error vectors are generated with a standard deviation $\sigma_\mathrm{pos}$. These vectors are low-pass filtered using a moving average window to reflect the temporal correlation typically observed in localization errors.
The filtered error vectors are then added to the localization data, thereby invalidating the assumed phase model and leading to a predicted degradation in SAR image quality.

\begin{figure}[tb]
	\centering
	\includegraphics[]{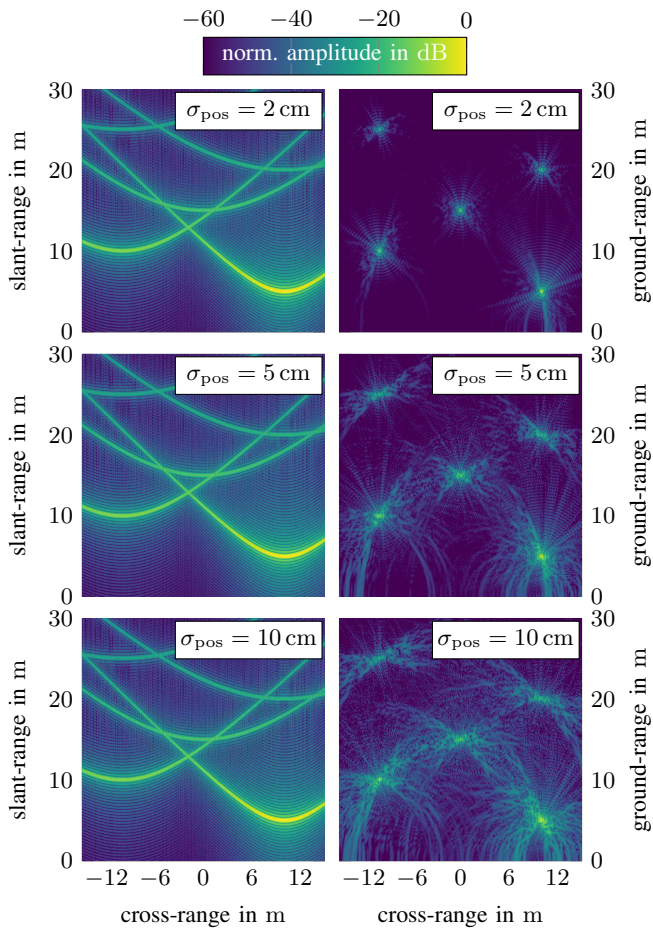}
	\caption{Simulation results of the scene in Fig.~\ref{fig:sim_scene} including range compression (left) and SAR image (right) for different localization errors.}
	\label{fig:sim_loc}
\end{figure}

Simulation results for varying levels of localization inaccuracy are presented in Fig.~\ref{fig:sim_loc}.
In these simulations, errors in estimating the sidelink path are incorporated into the computation of the expected phase in~(\ref{eq:signal_bp_expectied_phase}).
As a result, the range-compressed signals remain unaffected. 
However, the SAR image quality deteriorates due to the incorrect phase model utilized in the backprojection algorithm.

\section{Experimental Verification}
\label{sec:measurements}
Experimental measurements are conducted to assess both the coherency and overall performance of the proposed system concept.
The evaluation begins with an idealized bistatic radar setup to validate the coherency achieved by the concept. 
Subsequently, the complexity of the experiments is increased: first through a static measurement scenario, followed by a UAV-based bistatic SAR demonstration. 
In the latter, monostatic, bistatic, and combined multistatic SAR images are produced to demonstrate the system’s capabilities in an outdoor scenario.

\subsection{Coherency Analysis}
\label{subsec:meas_lab}
In the initial bistatic radar experiment, both the sidelink and radar signal paths are emulated using coaxial cables with lengths $\SI{5}{\meter}$ for the sidelink and $\SI{10}{\meter}$ for the radar signal.
The configuration parameters for the transmitting and receiving radar nodes are listed in Table~\ref{tab:radar_parameters}.
The received signals are sampled at the ADC with a rate of $f_\mathrm{s,ADC}$\,$=$\,$\SI[per-mode = symbol]{4.096}{\GS\per\second}$, digitally downconverted using a carrier at frequency $f_\mathrm{c}$, and subsequently decimated by a factor of $4$.
This results in an effective sampling rate of $f_\mathrm{s}$\,$=$\,$\SI[per-mode = symbol]{1.024}{\GS\per\second}$.

Both the sampling and carrier frequencies are derived from a reference clock with a frequency stability of $\SI{10}{\ppb}$~\cite{ReferenceClk2024}, which is well within the tolerance limits discussed in Section~V-\ref{subsec:errors_sfo}.
Under these conditions, the expected maximum SFO and CFO are  $f_\mathrm{SFO}$\,$=$\,$\SI{12}{\hertz}$ and $f_\mathrm{CFO}$\,$=$\,$\SI{41}{\hertz}$, respectively.
These values remain far below the maximum tolerable thresholds established in Sections~V-\ref{subsec:errors_sfo} and~\ref{subsec:errors_cfo}, suggesting that no significant performance degradation is expected due to SFO or CFO.
Nonetheless, slow-time effects such as CPE and TO must still be taken into account.

\begin{table}
\begin{center}
\caption{Radar Parameters of the Measurement System}
\label{tab:radar_parameters}
\begin{tabular}{| l | c | r |}
\hline
Parameter & Symbol & Value\\
\hline
carrier frequency & $f_\mathrm{c}$ & $\SI{1.2}{\giga\hertz}$\\
number of subcarriers & $N$ & $4096$\\
subcarrier spacing & $\Delta f$ & $\SI{100}{\kilo\hertz}$\\
bandwidth & $B$ & $\SI{409.6}{\mega\hertz}$\\
symbol duration & $T$ & $\SI{12.5}{\micro\second}$\\
cyclic prefix duration & $T_\mathrm{cp}$ & $\SI{3.125}{\micro\second}$\\
measurement rate & $f_\mathrm{prf}$ & $\SI{100}{\hertz}$\\
duty cycle & $\gamma$ & $\SI{0.156}{\percent}$\\
mean data rate & $C$ & $\SI[per-mode = symbol]{44.8}{\mega\bit\per\second}$\\
\hline
DAC\,/\,ADC sampling rate & $f_\mathrm{s}$ & $\SI[per-mode = symbol]{1.024}{\GS\per\second}$\\
DAC\,/\,ADC resolution & $\nu$ & $\SI{14}{\bit}$\\
\hline 
\end{tabular}
\end{center}
\end{table}

\begin{figure}[b]
	\centering
	\includegraphics[width=86mm,height=50mm]{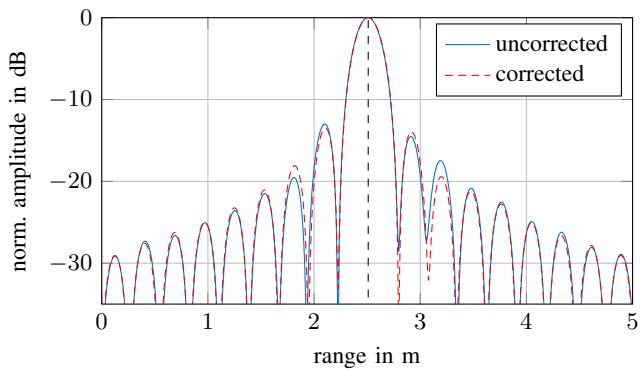}
	\caption{Range-compressed measured data of a cable-based measurement before and after the correction by applying (\ref{eq:signal_ofdm_rx_spectral_division_sl}) with ground truth indicated by a dashed line.}
	\label{fig:meas_lab_rc}
\end{figure}

Fig.~\ref{fig:meas_lab_rc} presents the range-compressed signals for both uncorrected and corrected OFDM radar measurements.
The correction is performed using sidelink-based calibration as described in~(\ref{eq:signal_ofdm_rx_spectral_division_sl}).
Visually, the range-compressed signals appear largely similar in both cases.

\begin{figure}[tb]
	\centering
	\includegraphics[width=86mm,height=50mm]{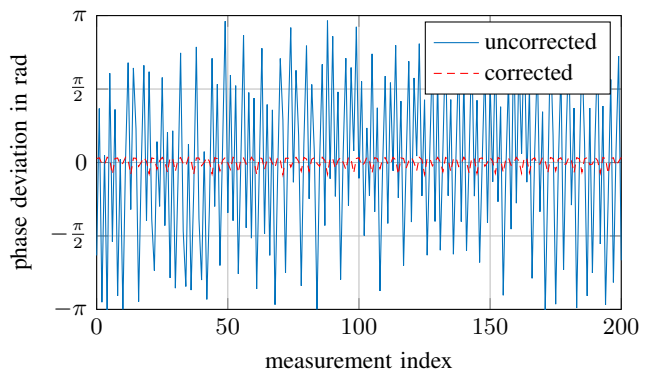}
	\caption{Phase distribution at the target peak in Fig.~\ref{fig:meas_lab_rc} before and after the correction by applying (\ref{eq:signal_ofdm_rx_spectral_division_sl}).}
	\label{fig:meas_lab_phase}
\end{figure}

However, for SAR imaging, phase information is critical.
To assess the impact of the correction step on phase stability, the phase at the target peak is extracted for each individual measurement.
A total of 200 consecutive measurements are analyzed.
In a static measurement scenario, the phase is expected to remain constant over time.
Fig.~\ref{fig:meas_lab_phase} illustrates the resulting phase deviations.

In the absence of correction, the phase varies within the interval $-\pi$ to $\pi$, primarily due to slow-time effects such as CPE and TO, as discussed in Sections~V-\ref{subsec:errors_cpe} and~\ref{subsec:errors_to}.
By contrast, the phase deviations in the corrected measurements are significantly reduced.

To quantify the stability of the phase over multiple measurements, the coherence factor is used as a metric~\cite{Hisatsu2020}. 
It is defined as
\begin{equation}
\label{eq:coherence_factor}
\Gamma_\mathrm{cf} 
= \frac{\left| \sum_{m=0}^{M-1} S_{\mathrm{B},m} (R_m) \right|^2}{M \cdot \sum_{m=0}^{M-1} \left| S_{\mathrm{B},m} (R_m) \right|^2}\,.
\end{equation}
In this context, $M$ denotes the total number of radar acquisitions, as introduced in~(\ref{eq:signal_bp_expectied_phase}). 
The coherence factor compares the coherent sum, as used in the backprojection algorithm in~(\ref{eq:signal_bp}), to the incoherent sum, which corresponds to the theoretical maximum amplitude. 
The interpretation of $\Gamma_\mathrm{cf}$ is as follows:
$\Gamma_\mathrm{cf}$\,$=$\,$0$ indicates destructive summation, $\Gamma_\mathrm{cf}$\,$=$\,$\frac{1}{M}$ corresponds to incoherent summation, and $\Gamma_\mathrm{cf}$\,$=$\,$1$ reflects fully coherent summation. 

For the bistatic measurements, the coherence factor is computed as $\Gamma_\mathrm{cf}$\,$=$\,$0.0034$ in the uncorrected case and $\Gamma_\mathrm{cf}$\,$=$\,$0.9860$ after applying the proposed correction. 
The reference value for incoherent summation, given by $\frac{1}{M}$\,$=$\,$0.005$ for $M$\,$=$\,$200$ measurements, confirms that the uncorrected signal is incoherent.
In contrast, the corrected result demonstrates high coherency, indicating that the processing approach detailed in Section~\ref{sec:processing} successfully preserves phase integrity with negligible loss compared to the ideal case. 

\subsection{Static Radar Measurement}
\label{subsec:meas_static}

\begin{figure}[tb]
	\centering
	\includegraphics[]{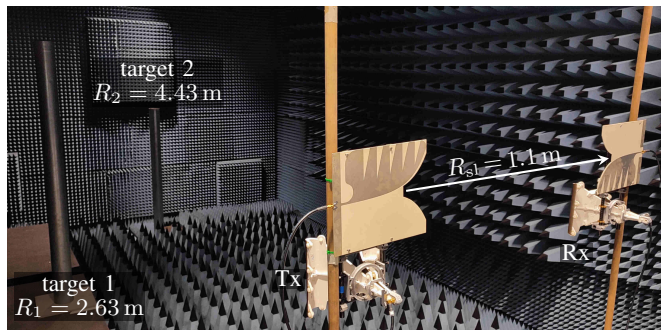}
	\caption{Photograph of the static measurement setup in the anechoic chamber consisting of a two-target scenario with ranges indicated in the graphic.}
	\label{fig:meas_static_setup}
\end{figure}

To further validate the proposed concept, radar measurements were conducted in an anechoic chamber in the context of a two-target scenario.
The experimental setup is shown in Fig.~\ref{fig:meas_static_setup}, with the geometric ranges annotated in the graphic.
Two metallic cylinders, each with a diameter of $\SI{15}{\centi\meter}$, are positioned within the scene.
The same antenna types utilized in the UAV-based system described in Section~\ref{sec:system} are employed for both the sidelink and radar signal paths.
System parameters are consistent with those listed in Table~\ref{tab:radar_parameters}.

To isolate the influence of coherency from geometric effects, the same physical configuration is used for both monostatic and bistatic measurements.
This is achieved by splitting the received signal.
In the monostatic case, the coherently received signal is sampled and recorded directly at the transmitting node.
The bistatic receiver operates independently and is not hardware-coupled to the monostatic system.
Nevertheless, coherency is established through the signal processing approach described in Section~\ref{sec:processing}.
The range offset $\hat{R}_\mathrm{cal}$ is determined via a calibration measurement, while the sidelink distance $R_\mathrm{sl}$ is calculated as the geometric distance between the phase centers of the sidelink antennas.

\begin{figure}[tb]
	\centering
	\includegraphics[width=86mm,height=50mm]{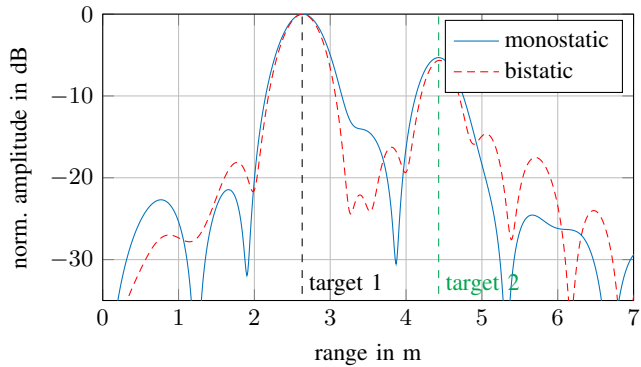}
	\caption{Range-compressed measured data of the monostatic and bistatic static measurement with ground truth indicated by dashed lines.}
	\label{fig:meas_static_rc}
\end{figure}

Fig.~\ref{fig:meas_static_rc} presents the range-compressed signals obtained after applying the processing steps outlined in Section~\ref{sec:processing}.
The displayed signal represents the coherent summation of range-compressed data from a sequence of $200$ individual measurements.
For the bistatic case, calibration is performed utilizing the sidelink signal, as described in~(\ref{eq:signal_ofdm_rx_spectral_division_sl}).
To reduce sidelobe levels, a Hann window is applied prior to compression.
In both the monostatic and bistatic results, the two targets are clearly identifiable as distinct peaks in the range-compressed data.

\begin{figure}[tb]
	\centering
	\includegraphics[width=86mm,height=50mm]{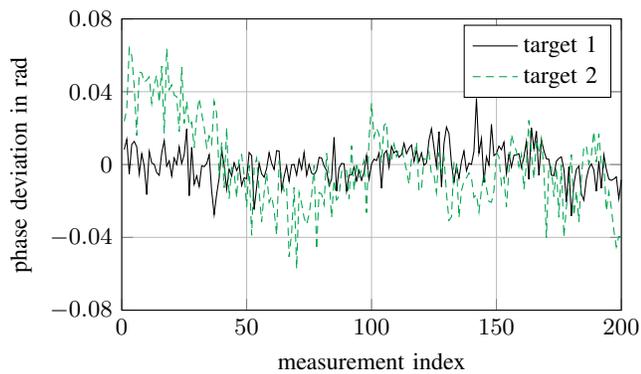}
	\caption{Phase distribution at the target peaks of the bistatic range-compressed data in Fig.~\ref{fig:meas_static_rc} for both measured targets.}
	\label{fig:meas_static_phase}
\end{figure}

As in the previous subsection, the phase stability at the target peaks is evaluated across the slow-time index $m$.
Fig.~\ref{fig:meas_static_phase} displays the phase deviations for both targets over a sequence of 200 measurements.
The target exhibiting a higher amplitude in the range-compressed data demonstrates smaller phase fluctuations compared to the second target.
The coherence factors are computed with~(\ref{eq:coherence_factor}) as $\Gamma_\mathrm{cf,1}$\,$=$\,$0.9999$ for the first target and $\Gamma_\mathrm{cf,2}$\,$=$\,$0.9995$ for the second. 
These results confirm that the proposed system and synchronization concept enable coherent processing of multistatic measurements.
This includes both slow-time coherency, which is essential for multistatic SAR imaging, and absolute coherency, which further allows for the coherent combination of monostatic and bistatic SAR images.

\subsection{UAV-Based Bistatic SAR Measurement}
\label{subsec:meas_uav}
To demonstrate the full capability of the system for coherent multistatic SAR imaging, a UAV-based experiment is conducted.
Two instances of the radar system described in Section~\ref{sec:system} are deployed, enabling simultaneous monostatic and bistatic SAR imaging within a single flight pass.

\begin{table}
\begin{center}
\caption{UAV-Based Measurement Geometry}
\label{tab:meas_geometry_uav}
\begin{tabular}{| l | c | r |}
\hline
Parameter & Symbol & Value\\
\hline
flight altitude above ground & $h$ & $4.5$\,$-$\,$\SI{5}{\meter}$\\
ground-range distance & $r_\mathrm{g}$ & $\SI{11.5}{\meter}$\\
trajectory length & $L_\mathrm{sa}$ & $\SI{3.4}{\meter}$\\
UAV distance & $R_\mathrm{sl}$ & $\SI{4.7}{\meter}$\\
depression angle & $\alpha$ & $\SI{45}{\degree}$\\
platform speed & $v$ & $\SI[per-mode = symbol]{0.5}{\meter\per\second}$\\
measurement rate & $f_\mathrm{prf}$ & $\SI{20}{\hertz}$\\
duty cycle & $\gamma$ & $\SI{0.031}{\percent}$\\
mean data rate & $C$ & $\SI[per-mode = symbol]{9.0}{\mega\bit\per\second}$\\
\hline
theoretical ground-range resolution & $\Delta r_\mathrm{g}$ & $\SI{58}{\centi\meter}$\\
theoretical cross-range resolution & $\Delta r_\mathrm{c}$ & $\SI{25}{\centi\meter}$\\
\hline 
\end{tabular}
\end{center}
\end{table}

The radar parameters are consistent with those utilized in the previous sections and are listed in Table~\ref{tab:radar_parameters}.
The indicated resolutions are $\SI{3}{\decibel}$\,-\,resolutions. 
An overview of the geometric configuration for this experiment is provided in Table~\ref{tab:meas_geometry_uav}.
The measurement scenario features an extended target placed in an environment with significant expected clutter.
The UAVs operate in a tandem formation, i.e., flying sequentially at the same altitude along a straight flight path.
A side-looking configuration is employed, with a depression angle of $45^\circ$, and the sidelink signal is transmitted by employing dedicated antennas, as illustrated in Fig.~\ref{fig:photo_uav_system}.

\begin{figure}[tb]
	\centering
	\includegraphics[]{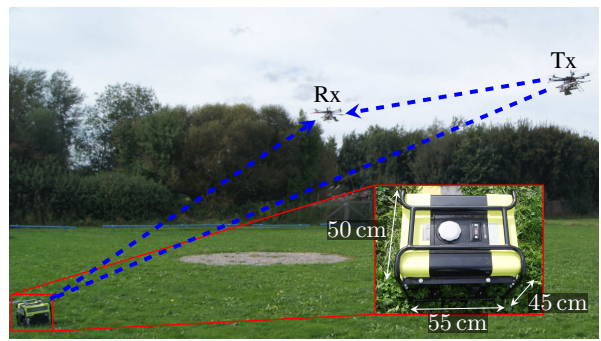}
	\caption{Photograph of the UAVs and the measured setup during the bistatic SAR measurement.}
	\label{fig:meas_uav_photo}
\end{figure}

\begin{figure}[tb]
	\centering
	\includegraphics[width=86mm,height=54mm]{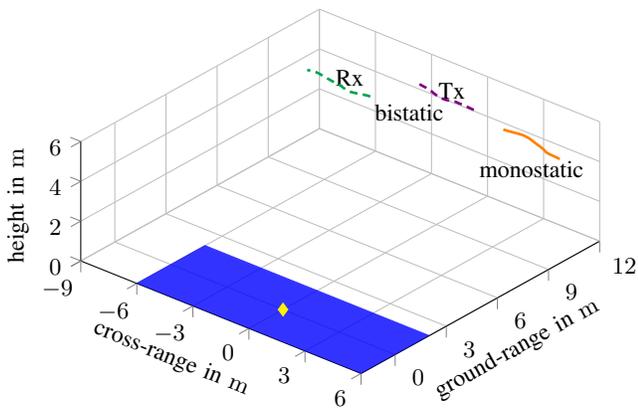}
	\caption{Visualization of the measurement geometry for the monostatic~(\protect\blacklinesolid) and bistatic~(\protect\blacklinedashed) acquisition with pixel grid (\protect\bluebox) and target position (\protect\yellowmark) indicated.}
	\label{fig:meas_uav_geom}
\end{figure}

Fig.~\ref{fig:meas_uav_photo} shows an image of the two UAVs in operation during the SAR measurement campaign.
A power generator is used as the primary target within the scene.
The layout of the measurement area and the UAV flight paths are depicted in Fig.~\ref{fig:meas_uav_geom}.
SAR imaging is performed over a region measuring $\SI{12}{\meter}$ in the cross-range direction and $\SI{4}{\meter}$ in the ground-range direction.
To enhance the imaging diversity and extend the effective aperture, distinct segments of the same flight trajectory are selected for the monostatic and bistatic evaluations.
Given the verified slow-time coherency from Sections~VI-\ref{subsec:meas_lab} and~\ref{subsec:meas_static}, these SAR images can be coherently combined despite being acquired from different viewing angles.

Following the acquisition, the recorded RF signal samples stored onboard each UAV are processed by employing the procedure outlined in Section~\ref{sec:processing}.
Consistent with the approach in the previous subsection, a Hann window is applied to the range-compressed data to suppress sidelobes.

\begin{figure}[tb]
	\centering
	\includegraphics[]{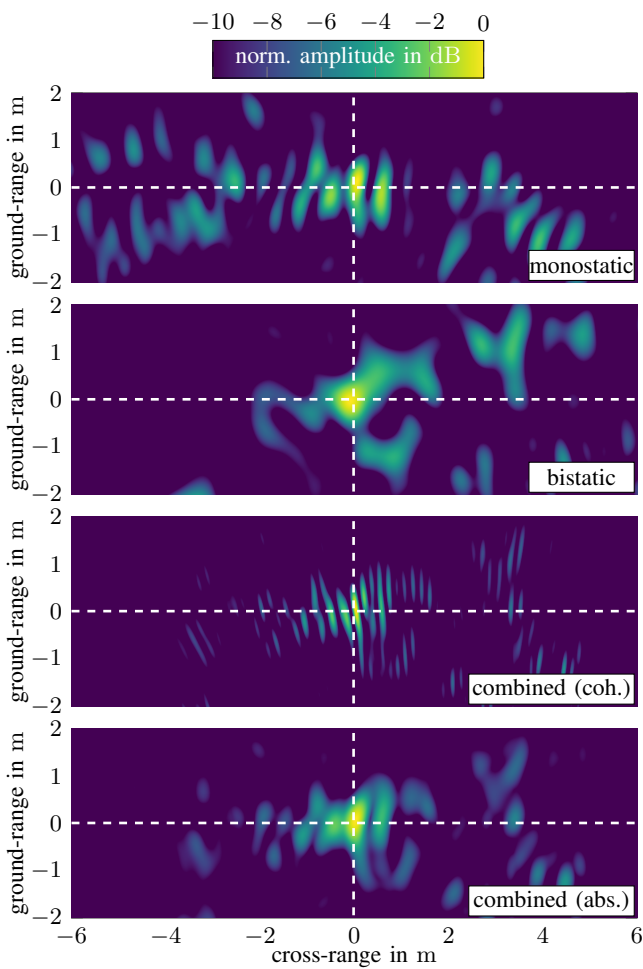}
	\caption{Comparison of SAR images of the measured scene in Fig.~\ref{fig:meas_uav_photo} based on the geometry illsutrated in Fig.~\ref{fig:meas_uav_geom}.}
	\label{fig:meas_uav_sar_combined}
\end{figure}

Fig.~\ref{fig:meas_uav_sar_combined} presents the resulting SAR images.
In both the monostatic and bistatic images, the target is clearly resolved and well-focused.
However, clutter patterns differ between the two due to variations in observation and scattering angles.
This diversity is leveraged to reduce clutter through image combination. 

One approach to produce a combined multistatic SAR image is the coherent summation of the normalized monostatic and bistatic SAR images.
Each image is first scaled to its respective peak value before combining.
The resulting multistatic SAR image is shown as the third panel from the top in Fig.~\ref{fig:meas_uav_sar_combined}. 
It demonstrates the effects of the extended synthetic aperture depicted in Fig.~\ref{fig:meas_uav_geom}, which includes a gap between the monostatic and the bistatic acquisitions.
As expected, this leads to enhanced resolution but introduces pronounced sidelobes near the main target peak, consistent with the theoretical findings in~\cite{Goodman2003}.
Compared to the individual images, the fused result exhibits improved clutter suppression, attributable to the angular diversity between the monostatic and bistatic perspectives.

As an alternative to coherent fusion, the SAR images can also be combined by summing their absolute pixel magnitudes.
The result of this non-coherent combination is presented in the fourth panel of Fig.~\ref{fig:meas_uav_sar_combined}.
Since this method does not preserve phase information, it does not yield an improvement in cross-range resolution.
Nevertheless, the clutter suppression remains comparable to that observed in the coherently combined image, benefiting similarly from the diversity in observation and scattering geometry. 

\begin{figure}[tb]
	\centering
	\includegraphics[]{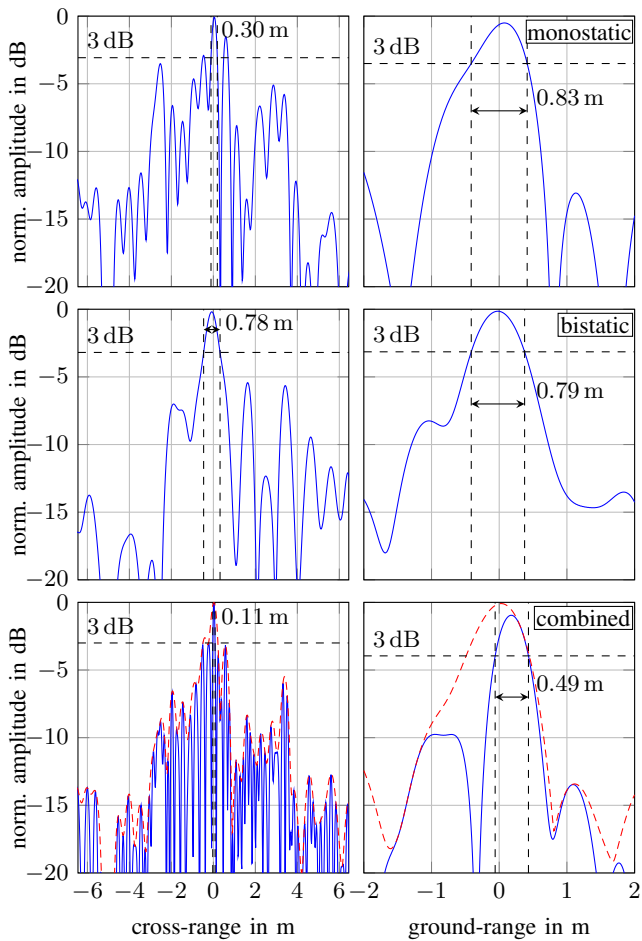}
	\caption{Cross-range and ground-range cuts through the SAR images shown in Fig.~\ref{fig:meas_uav_sar_combined} with indicated $\SI{3}{\decibel}$\,-\,resolutions. The bottom row presents both coherently~(\protect\blacklinesolid) and absolutely~(\protect\blacklinedashed) combined SAR image profiles.}
	\label{fig:meas_uav_sar_cuts}
\end{figure}

To further assess the imaging performance enabled by the coherent processing approach, cross-sectional profiles through the SAR images are shown in Fig.~\ref{fig:meas_uav_sar_cuts}.
For a fair comparison between the monostatic, bistatic, and combined results, the peak positions in the individual SAR images are identified, and the geometric midpoint between these maxima is selected as the reference for evaluation.
The locations of the cross-range and ground-range cuts are indicated in Fig.~\ref{fig:meas_uav_sar_combined}.
The resulting profiles include markers for the $\SI{3}{\decibel}$\,-\,widths to illustrate the achieved resolution.
While the cross-range and ground-range resolutions are close to the expected values, slight deviations are observed.
These discrepancies are likely attributed to the physical extent of the imaged object, which may slightly violate the point-target assumption used in the resolution estimates.

In the combined SAR image, the improved cross-range resolution is evident, along with sidelobes near the target peak resulting from the gap in the synthetic aperture.
A cross-range resolution of $\SI{11}{\centi\meter}$ is achieved.
In the ground-range direction, resolution also improves to $\SI{49}{\centi\meter}$.
At wider azimuth angles, the ground-range resolution becomes increasingly influenced by the synthetic aperture, rather than being solely determined by the radar bandwidth.
The profiles extracted from the absolutely combined SAR image effectively serve as an envelope of the coherently fused result, as the phase information is not retained.

\section{Conclusion}
\label{sec:conclusion}
Recent advancements in high-speed ADCs and DACs have made it practical to implement fully digital radar architectures. 
These systems can directly sample RF signals, eliminating the need for analog downconversion. 
As a result, coherency at the RF level can be maintained through digital post-processing, significantly reducing the need for stringent synchronization of reference signals. 
Moreover, digital radar systems offer considerable flexibility in waveform design and signal processing.

This work presents a UAV-based digital radar platform capable of performing simultaneous monostatic and multistatic SAR measurements in the L-band. 
Leveraging the flexibility of digital radar, a synchronization approach is introduced that substantially relaxes the precision requirements  by several orders of magnitude compared to conventional techniques. 
The radar signal is transmitted through both the radar and sidelink paths, ensuring full coherency in both slow-time and fast-time domains. 
To reduce the average data rate, the system employs a known trigger sequence to selectively store only relevant radar data.

The passive configuration of the multistatic receive nodes supports scalability, enabling deployment across large UAV swarms. 
A full signal processing based on OFDM is detailed for both monostatic and bistatic SAR imaging modes. 
Potential errors introduced by the autonomous operation of individual nodes are thoroughly analyzed. 
These errors are inherently mitigated within the proposed signal processing.

System performance is validated through measurements, including a UAV-based bistatic SAR experiment. 
The results demonstrate that the proposed synchronization scheme successfully preserves coherency in slow-time and fast-time. 
Monostatic and bistatic SAR images as well as a combined multistatic SAR images are generated, exhibiting improved cross-range resolution in the coherently combined case.

\bibliographystyle{IEEEtran}

\bibliography{2024-journal-bistatic_uav_pmcw_references.bib}

\end{document}